\documentclass[a4paper]{jpconf}

\usepackage{graphicx}
\usepackage{amsmath}
\usepackage{amssymb}

\begin{document}

\title{Coherence of Noisy Oscillators with Delayed Feedback Inducing Multistability}

\author{Anastasiya V Pimenova$^1$ and Denis S Goldobin$^{1,2,3}$}
\address{$^1$Institute of Continuous Media Mechanics, UB RAS,
             Perm 614013, Russia}
\address{$^2$Department of Theoretical Physics, Perm State University,
             Perm 614990, Russia}
\address{$^3$Department of Mathematics, University of Leicester,
             Leicester LE1 7RH, UK}
\ead{Anastasiya.Pimenova@gmail.com, Denis.Goldobin@gmail.com}

\begin{abstract}
For self-sustained oscillators subject to noise the coherence, understood as a constancy of the instantaneous oscillation frequency, is one of the primary characteristics. The delayed feedback has been previously revealed to be an efficient tool for controlling coherence of noise-driven self-sustained oscillator. The effect of the delayed feedback control on coherence is stronger for a longer delay time. Meanwhile, the instantaneous frequency of a noise-free oscillator can exhibit multistability for long delay time. The impact of the delay-feedback-induced multistability on the oscillation coherence, measured by the phase diffusion constant, of a noisy oscillator is studied in this work both numerically and analytically.
\end{abstract}

\section{Introduction}
Delayed feedback was found to be a highly efficient tool for controlling the coherence of noisy oscillators~\cite{Goldobin-Rosenblum-Pikovsky-2003a,Goldobin-Rosenblum-Pikovsky-2003b,Boccaletti-Allaria-Meucci-2004,Goldobin-2014}.  Even weak feedback with sufficiently long delay time can diminish or enhance the phase diffusion constant, quantitative measure of coherence, by one order of magnitude and even more.  The theory of the delay feedback control of the phase diffusion was developed in~\cite{Goldobin-Rosenblum-Pikovsky-2003a,Goldobin-Rosenblum-Pikovsky-2003b,Goldobin-2014}.

However, for long delay times and vanishing noise, the multistability of the mean frequency can occur. In the presence of multistability, noise results in intermittent switchings between states with different mean frequencies and, thus, the phase diffusion is contributed not only by fluctuations around the mean linear growth of the phase but also by the alternation of `local' mean growth rates.  For weak noise---which is a physically relevant situation---it turns-out to be possible to find natural variable in terms of which the switching between two stable phase growth rates becomes a perfect telegraph process. In~\cite{DHuys-Jungling-Kinzel-2014}, frequency multistability and noise-induced switchings were studied for extremely long delay times, although without consideration of coherence and phase diffusion.

In this work, on the basis of the `telegraphness' property, we construct the analytical theory of the effect of delayed feedback on the phase diffusion in the presence of multistability. In agreement with the results of numerical simulation we analytically derive that the phase diffusion constant has giant peaks close to the points where the residence times in two states are equal.

\section{Phase Reduction}
The dynamics of a limit-cycle oscillator subject to weak action can be described within the framework of the phase reduction, where the system state in determined solely by the oscillation phase~\cite{Winfree-1967, Kuramoto-2003}. The phase reduction can be used as well for the case of weak noise, including a $\delta$-correlated one~\cite{Yoshimura-Arai-2008, Goldobin-etal-2010}. For nearly harmonic oscillators subject to white Gaussian noise and linear delayed feedback the phase equation reads~\cite{Goldobin-Rosenblum-Pikovsky-2003b, Kuramoto-2003}
\begin{equation}
\dot{\varphi}=\Omega_0+a\sin[\varphi(t-\tau)-\varphi(t)]+\varepsilon\xi(t)\,,
\label{eq-01}
\end{equation}
where $\Omega_0$ is the natural frequency of the oscillator, $a$ and $\tau$ are the strength and delay time of the feedback, respectively, $\varepsilon$ is the noise strength, $\xi(t)$ is the normalised $\delta$-correlated Gaussian noise:
\[
\langle\xi(t)\xi(t+t')\rangle=2\delta(t'),\qquad\langle\xi\rangle=0.
\]
Recently, a strong impact of weak anharmonicity for recursive delay feedback has been reported for the effects under our consideration~\cite{Goldobin-2011}; however, for a `simple' delay the harmonic approximation and model reduction~\eref{eq-01} remain well justified for many physical, chemical and biological systems ({\it e.g.}, see~\cite{Kuramoto-2003,Masoller-2002}).

Noise disturbs the linear growth of the phase, resulting in its diffusion according to the law
$\langle[\varphi(t)-\langle\varphi(t)\rangle]^2\rangle=Dt$ for $t\to\infty$. In the control-free case, $D=D_0=2\varepsilon^2$. The diffusion constant $D$ measures the coherence of oscillations; the smaller $D$ the better coherence. Within the framework of the linear-in-noise approximation, delayed feedback has been found to change the mean oscillation frequency $\Omega$ and the diffusion constant $D$ as follows~\cite{Goldobin-Rosenblum-Pikovsky-2003a, Goldobin-Rosenblum-Pikovsky-2003b, Goldobin-2011};
\begin{eqnarray}
&&\Omega=\Omega_0-a\sin{\Omega\tau}\,,
\label{eq-02}
\\
&&D=\frac{D_0}{(1+a\tau\cos{\Omega\tau})^2}\,.
\label{eq-03}
\end{eqnarray}

\begin{figure}[t]
\center{
\includegraphics[width=0.85\textwidth]{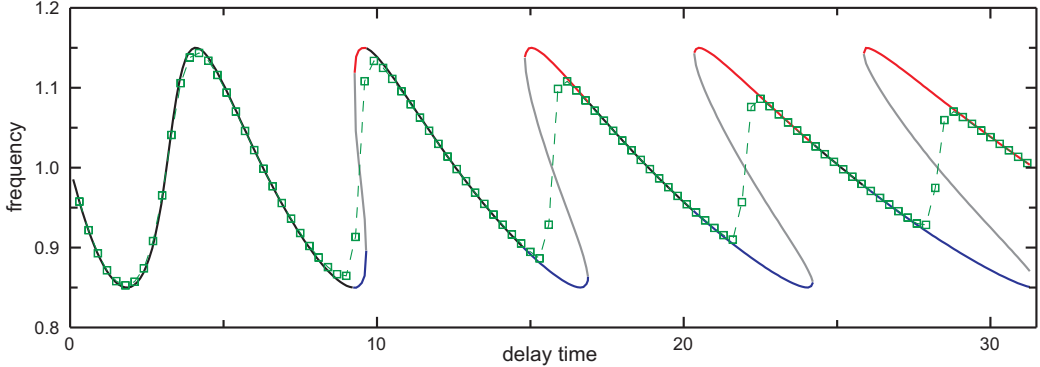}}
\caption{Mean oscillation frequency $\Omega$ {\it vs} delay time $\tau$ for $\Omega_0=1$ and $a=0.15$. Solid line: the noise-free case (solution to equation~\eref{eq-02}). Squares: numerical simulation for $\varepsilon^2=0.02$.}
\label{fig1}
\end{figure}

\section{`Telegraphness' of the instantaneous frequency in the presence of weak noise}
\subsection{Multistability of oscillation frequency for vanishing noise}
Let us first discuss the system dynamics for vanishing noise. In \fref{fig1} the solution of equation~\eref{eq-02} is plotted for the noise-free case with the solid line. One can notice that for large $\tau$ the frequency $\Omega$ is not always unique, which has already been noticed in~\cite{Goldobin-Rosenblum-Pikovsky-2003b, Niebur-Schuster-Kammen-1991}. The linear stability analysis for this system reveals that the linear growth solution $\varphi(t)=\Omega t$ is unstable when $a\tau\cos{\Omega\tau}<-1$. In~\cite{Goldobin-Rosenblum-Pikovsky-2003b}, the stability analysis can be found for the monotonous perturbations, the analysis for the oscillatory perturbations (to be published elsewhere) does not change this condition. If one moves along the curve in \fref{fig1} from the origin, this condition will be fulfilled on the segments of the line going from the minimal values of $\Omega$ towards the maximal ones between points with vertical inclination. These segments, say $\omega_0$, are plotted with gray and separate the branches of stable solutions which are plotted with blue, $\omega_1$, and red, $\omega_2$, in the domains of multistability.

Equation system~\eref{eq-02}--\eref{eq-03} is derived within the framework of the linear-in-noise approximation, which considers only small fluctuations around a state with constant phase growth rate $\Omega$; it does not handle hopping between several states with $\Omega=\omega_i$. Equation~\eref{eq-03} has to be considered as a `local' phase diffusion at the state with corresponding $\Omega$. The system dynamics is dominantly controlled by the delay term, which makes the events of transition between two stable states of the noise-free system non-local in time; for instance, $\dot{\varphi}$ can be kicked-out by noise from $\omega_1$, reach $\omega_2$ and then again return to $\omega_1$. Thus, the noise-driven dynamics of the system in the presence of multistability is a non-trivial problem requiring subtle consideration. For longer delay time $\tau$ there are domains with more than two locally stable regimes. As a first step, in this work we restrict our consideration to the case of multistability between two stable regimes; for $\Omega_0=1$ and $a=0.15$ (\fref{fig1}) this is the case as long as $\tau\lesssim 31.5$.

\begin{figure}[t]
\center{
\begin{tabular}{cc}
{\sf (a)}\hspace{-10pt}
\includegraphics[width=0.44\textwidth]{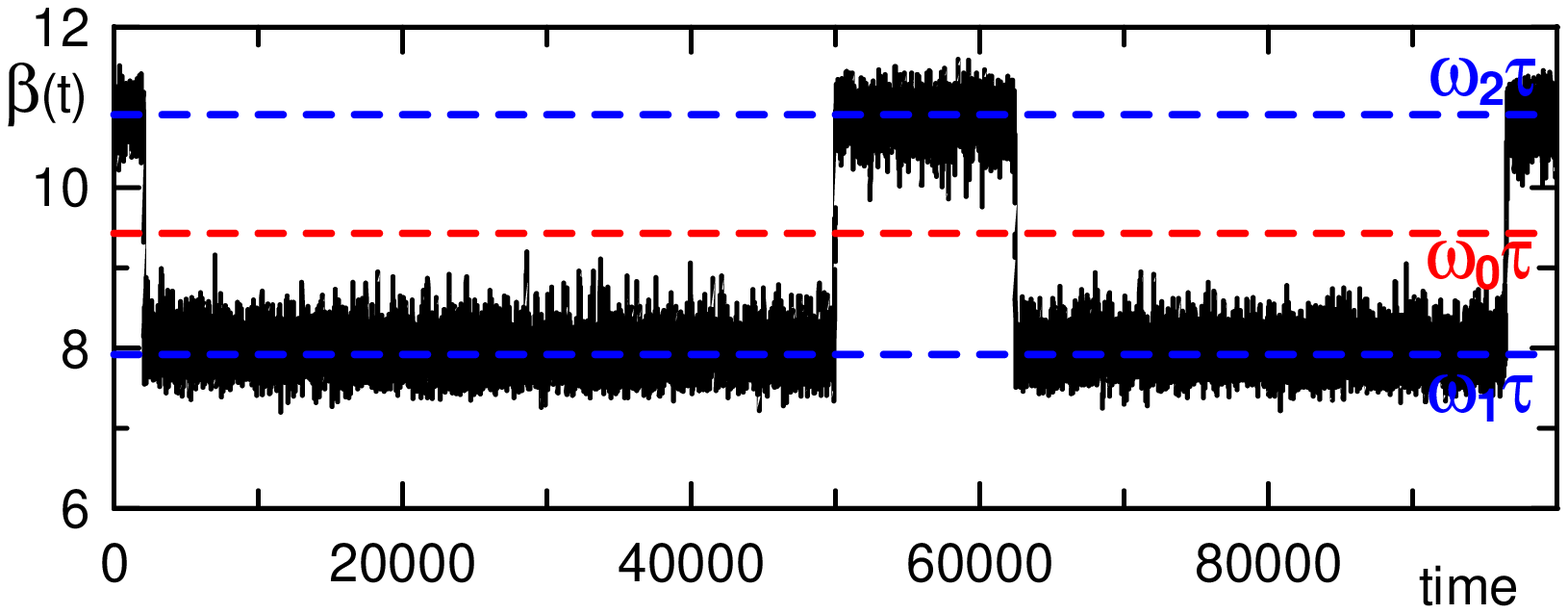}
&
{\sf (b)}\hspace{-10pt}
\includegraphics[width=0.44\textwidth]{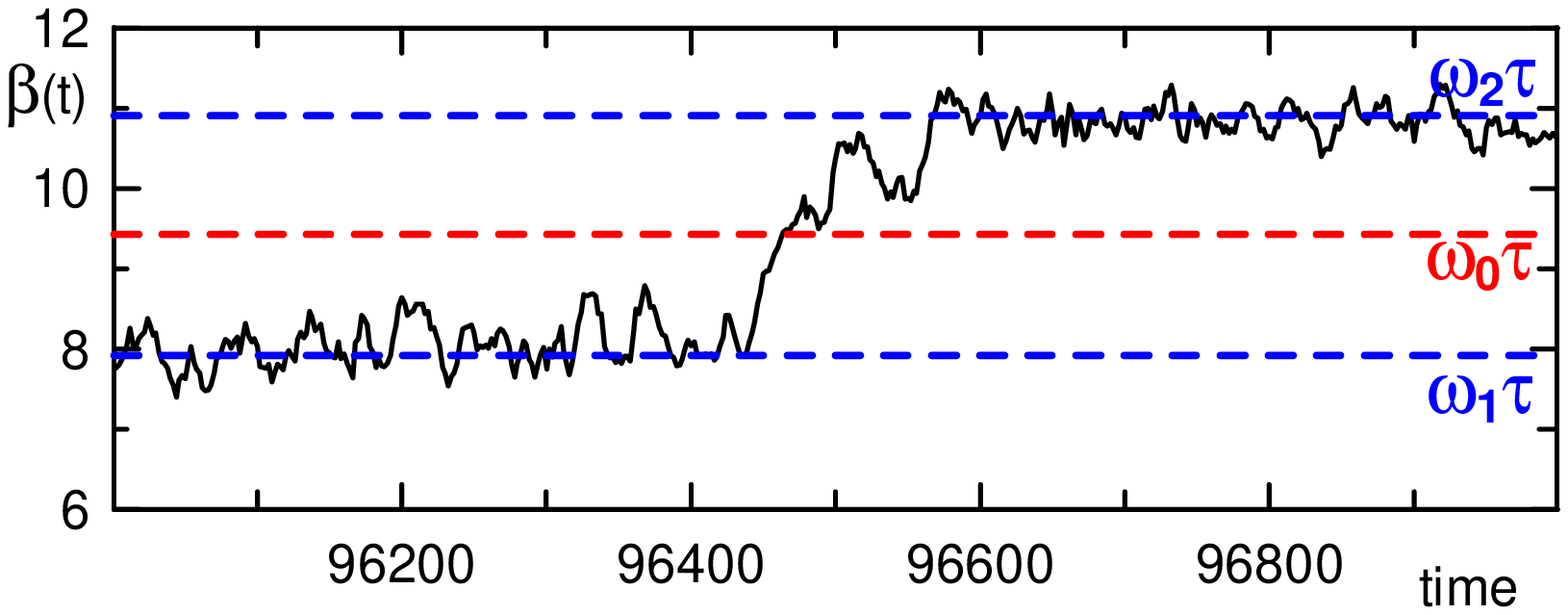}\\[15pt]
{\sf (c)}\hspace{-10pt}
\includegraphics[width=0.38\textwidth]{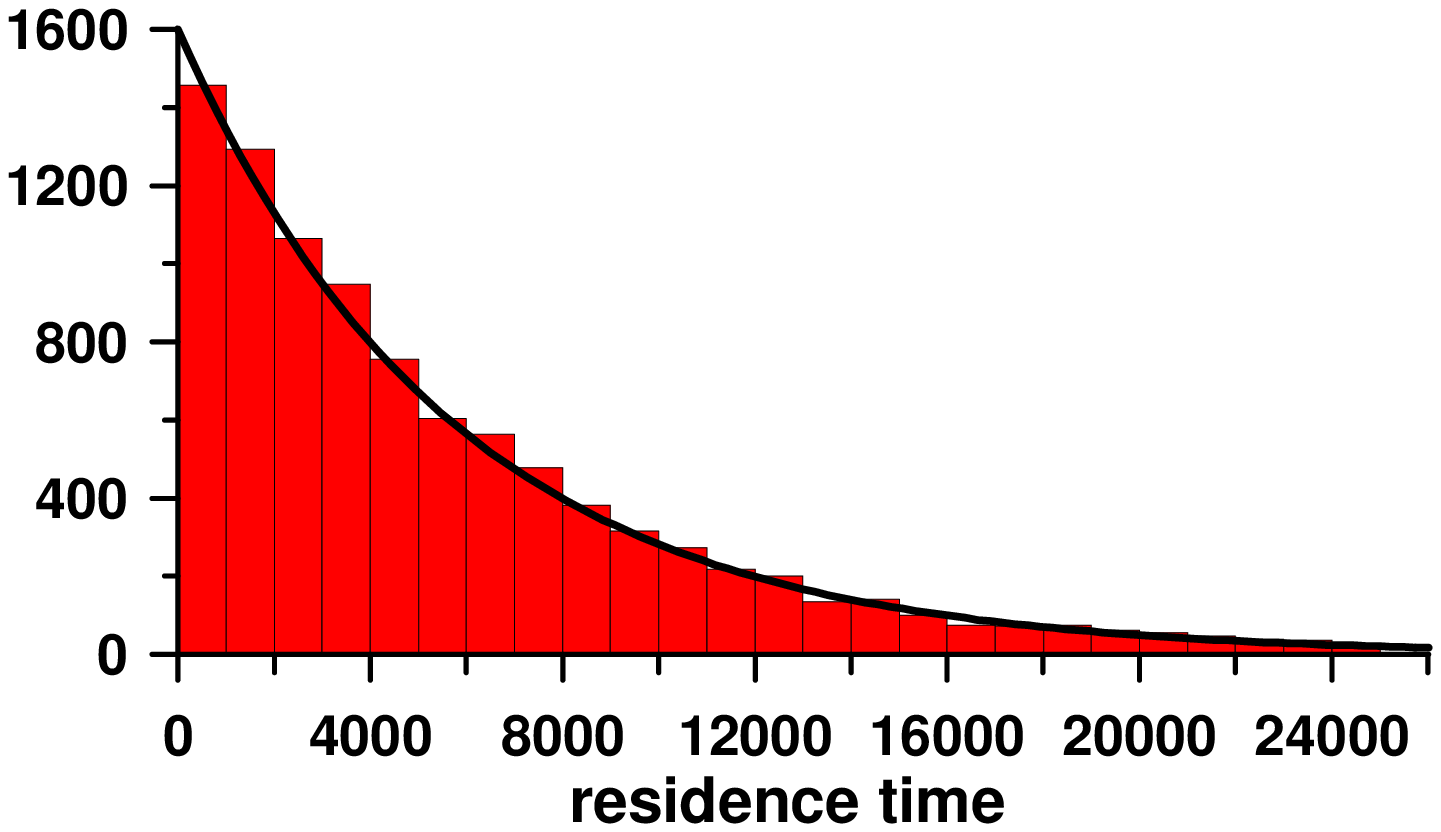}
&
{\sf (d)}\hspace{-10pt}
\includegraphics[width=0.38\textwidth]{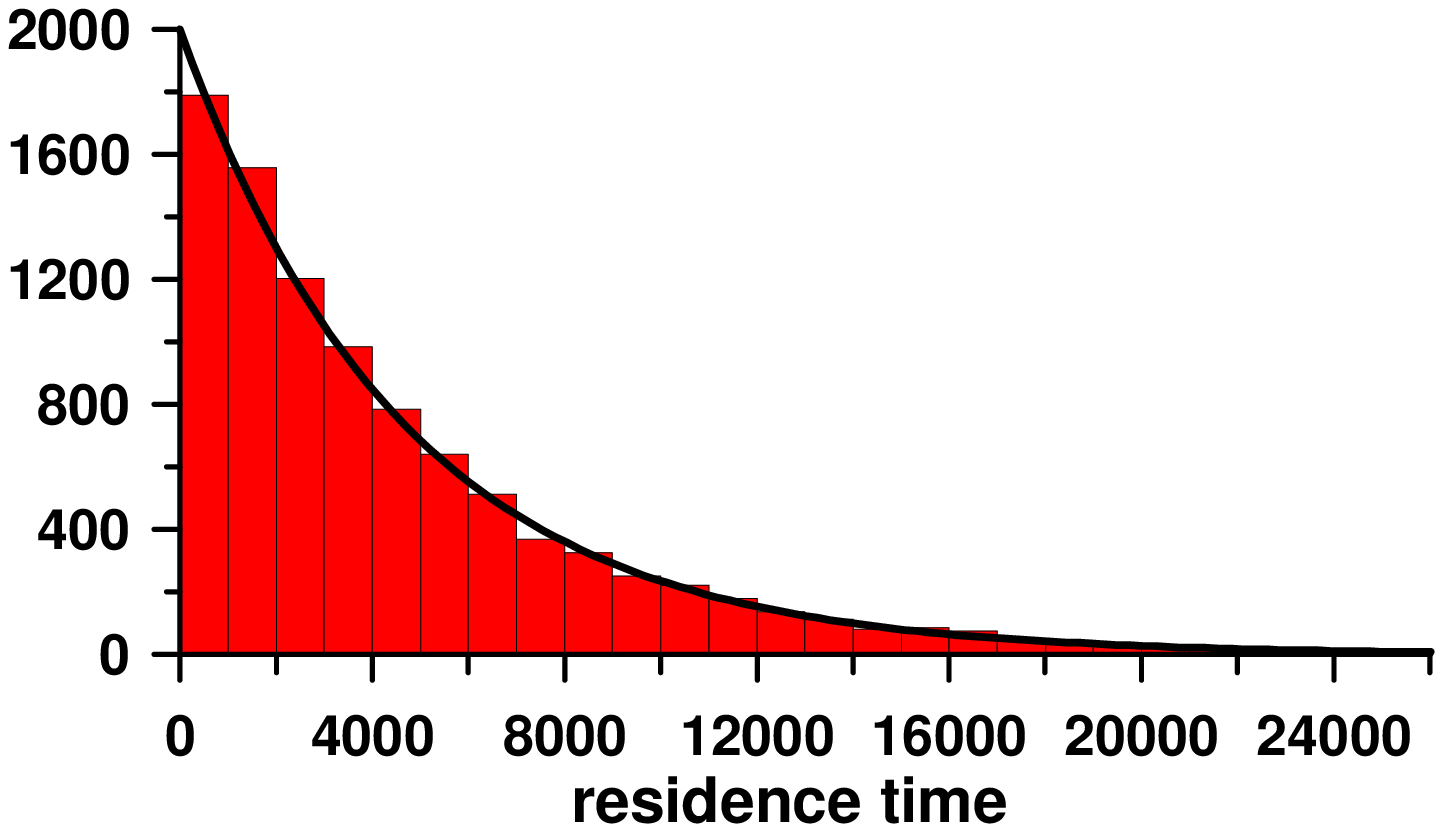}
\end{tabular}
}
\caption{Sample of the dynamics of the system with multistability in the presence of weak noise for $\varepsilon^2=0.02$, $\Omega_0=1$, $\tau=9.5$.
(a) and (b): dynamics of $\beta(t)=\varphi(t)-\varphi(t-\tau)$ in time. (c) and (d): the distribution of residence times for $\omega_1$ and $\omega_2$, respectively, for $9551$ switching events; the solid lines are exponential fitting of these distributions.}
\label{fig2}
\end{figure}

\subsection{Dynamics of the system subject to weak noise}
Let us rewrite equation~\eref{eq-01} in the form
\[
\omega(t)\equiv\dot{\varphi}=\Omega_0-a\sin\beta(t)+\varepsilon\xi(t)\,,
\qquad\dot{\beta}(t)=\omega(t)-\omega(t-\tau)\,,
\]
where $\beta\equiv\varphi(t)-\varphi(t-\tau)$ is the phase growth per delay time. This integral form of our dynamic system suggests that $\beta(t)$ can be a suitable `natural' variable to trace the switchings between two states. Indeed, in \fref{fig2} one can see that in terms of $\beta$ the system state is a `telegraph' signal; transitions between two states occur without immediate reversals---each reverse transition occurs as an event independent of the previous transition---the residence time distributions for two states are exponential. This `telegraphness' can be used to develop analytical theory of the dynamics of a system with multistability.

\section{Phase diffusion in the presence of multistability}
\subsection{Analytical theory}
The diffusion constant of an asymmetric telegraph process altering between two states with growth rates $\omega_1$ and $\omega_2$ and mean residence times $T_1$ and $T_2$, respectively, can be calculated;
\begin{equation}
D_\mathrm{tel}=\frac{2(\omega_2-\omega_1)^2T_1^2T_2^2}{(T_1+T_2)^3}\,.
\label{eq-04}
\end{equation}
The net phase diffusion is contributed by `telegraph' diffusion and local diffusions at states $\omega_{1,2}$;
\begin{equation}
D=D_\mathrm{tel}+\frac{T_1}{T_1+T_2}D_1+\frac{T_2}{T_1+T_2}D_2\,,
\label{eq-05}
\end{equation}
where $D_{1,2}=D_0/(1+a\tau\cos{\omega_{1,2}\tau})^2$ (equation~\eref{eq-03}).

\begin{figure}[t]
\center{
\includegraphics[width=0.91\textwidth]{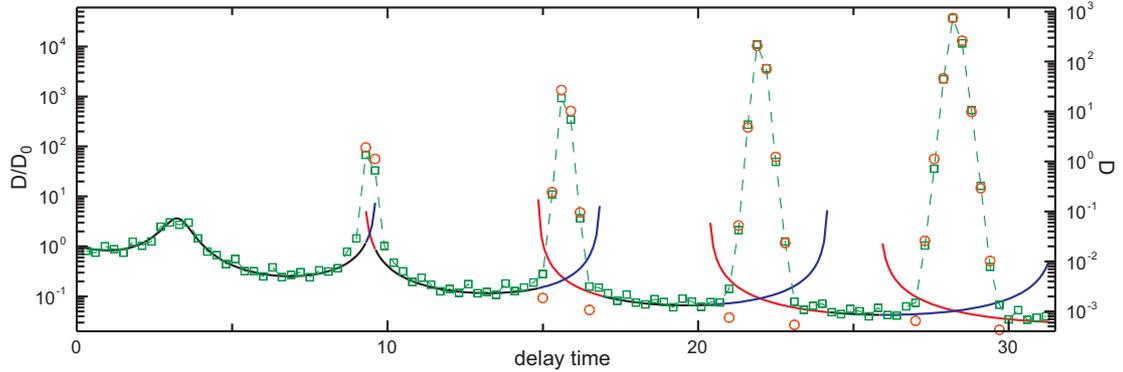}}
\caption{Phase diffusion constant $D$ {\it vs} delay time $\tau$ for $\Omega_0=1$, $a=0.15$ and $\varepsilon^2=0.02$. Solid lines: `local' diffusion at stable states. Green squares: numerical simulation. Orange circles: diffusion constant (equation~\eref{eq-04}) of a telegraph process with given $\omega_{1,2}$ and mean residence times $T_{1,2}$ taken from the results of numerical simulation.}
\label{fig3}
\end{figure}

\subsection{Phase diffusion on the edges of the multistability domains}
Singular behaviour is observed for the local phase diffusion at the edges of the multistability domains (\fref{fig3}). In \fref{fig3} the local phase diffusions $D_i$ are plotted with the solid line. The line is black for the domains with no multistability and coloured otherwise; the same colours mark corresponding regimes in \fref{fig1}. One can notice that at the edges of the multistability domains for the regime experiencing the tangential bifurcation at this edge the local phase diffusion~\eref{eq-03} tends to infinity. This divergence of local diffusion rises the question how the net diffusion behaves on the edges of the multistability domains. To answer this question, one can perform a normal form analysis of the system near the tangential bifurcation in the presence of weak noise and derive scaling properties for quantifiers of interest.

Let us consider a point of the tangential bifurcation where regimes $\omega_0$ and $\omega_2$ appear (annihilate), $\tau=\tau_{2\ast}$, and describe the noise-induced behaviour of the system for $0<\tau-\tau_{2\ast}\ll 1$. The normal form of the system dynamics near the bifurcation point reads $\dot{x}=-dU/dx+\varepsilon\xi(t)$ with $U(x)=-\alpha x+x^3/3$ and $\alpha=k_2(\tau-\tau_{2\ast})$, where $k_2$ is a positive number of order of magnitude of $1$. For this normal form the mean first passage time from the potential well can be calculated for weak noise;
\begin{equation}
T_2(\tau\approx\tau_{2\ast})\approx\frac{\pi}{2}[k_2(\tau-\tau_{2\ast})]^{-\frac{1}{2}}
 \exp\left(\frac{4}{3\varepsilon^2}[k_2(\tau-\tau_{2\ast})]^\frac{3}{2}\right).
\label{eq-06}
\end{equation}
Since $T_2\ll T_1$, the telegraph diffusion~\eref{eq-04} becomes $D_\mathrm{tel}(\tau\approx\tau_{2\ast})\approx2(\omega_2-\omega_1)^2(T_2^2/T_1^3)$, which tends to zero for $\tau\to\tau_{2\ast}$. Subtle analysis of equations~\eref{eq-02}--\eref{eq-03} yields the scaling law for the local diffusion $D_2\propto D_0/(\tau-\tau_{2\ast})$; it will be convenient to introduce constant $K_2$ as follows $D_2=(D_0/2)K_2\sqrt{k_2}/(\tau-\tau_{2\ast})$. The behaviour of $D_1$ is regular around $\tau=\tau_{2\ast}$. Hence, equation~\eref{eq-04} takes the form
\begin{equation}
D=D_\mathrm{tel}+D_1+\frac{K_2\varepsilon^2}{T_1(\tau-\tau_{2\ast})^{3/2}}
 e^{\frac{4}{3\varepsilon^2}[k_2(\tau-\tau_{2\ast})]^{3/2}}\,.
\label{eq-07}
\end{equation}
One can see that for $\tau-\tau_{2\ast}>\varepsilon$, where the weak noise approximation and our derivations are valid, the last term in~\eref{eq-07} rapidly decreases as $\tau$ tends to $\tau_{2\ast}$. Thus, both the telegraph diffusion and the contribution of the diverging local diffusion into the net diffusion vanish at the edge of the multistability domains, and the behaviour of the net diffusion is regular, as it can be also seen from the results of numerical simulation (see~\fref{fig3}).

\begin{figure}[t]
\center{
\includegraphics[width=0.85\textwidth]{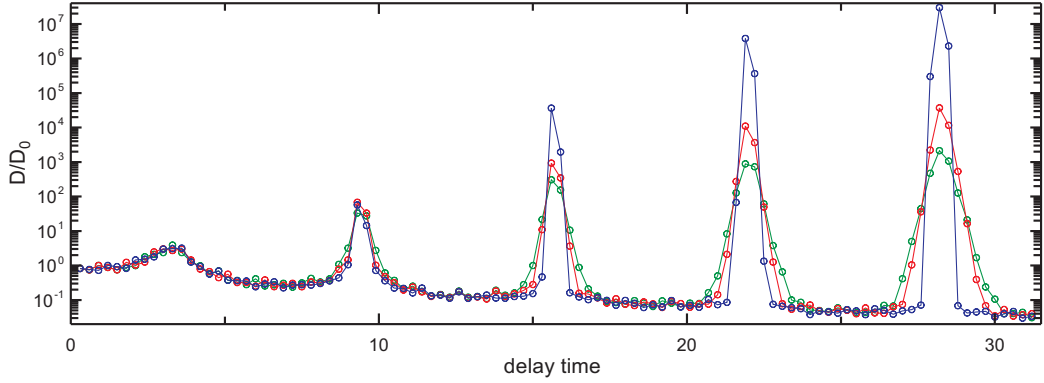}}
\caption{Phase diffusion constant $D$ {\it vs} delay time $\tau$: numerical simulation for $\Omega_0=1$, $a=0.15$ and $\varepsilon^2=0.01$ (blue), $0.02$ (red), $0.03$ (green).}
\label{fig4}
\end{figure}

\subsection{Behaviour of phase diffusion near the centres of the multistability domains}
Within a multistability domain the mean residence time $T_1$ decreases as $\tau$ increases and tends to zero at the right edge of the multistability domain, while $T_2$ increases starting from zero at the left edge of the domain. At certain value $\tau_\ast$ these times become equal, $T_1=T_2=T_\ast$. Let us consider the vicinity of this point. For small deviations $\delta\tau=\tau-\tau_\ast$, one can write
 $T_{1,2}\approx T_\ast\exp[(d\ln{T_{1,2}}/d\tau)\delta\tau]
 \equiv T_\ast\exp[\mp(q_{1,2}/\varepsilon^2)\delta\tau]$.
Notice, since the mean residence times depend on $\tau$ exponentially strong with small number $\varepsilon^2$ in the denominator of the argument of the exponential, we linearise not $T_i$ but $\ln{T_i}$ next to $\tau_\ast$. Coefficients $q_{1,2}$ are positive number of order of magnitude of $1$. Hence,
\begin{equation}
D(\tau\approx\tau_\ast)=\frac{(\omega_2-\omega_1)^2T_\ast}{4}
 \frac{\displaystyle\exp\left(\frac{q_2-q_1}{2\varepsilon^2}(\tau-\tau_\ast)\right)}
 {\displaystyle\cosh^3\left(\frac{q_1+q_2}{2\varepsilon^2}(\tau-\tau_\ast)\right)}
 +\frac{D_1+D_2}{2}\,.
\label{eq-08}
\end{equation}
The exponential in the nominator is a small correction to the cube of hyperbolic cosine in the denominator. One has to expect a strong peak near $T_1=T_2$ with height somewhat above $(\omega_2-\omega_1)^2T_\ast/4$. Notice, for the weak noise, the latter expression is exponentially large, as $T_\ast$ is exponentially large. Moreover, for vanishing noise the width of the peak $\propto\varepsilon^2$, while its height is exponentially large in $1/\varepsilon^2$, {\it i.e.}, the integral of this peak over $\tau$ diverges exponentially fast.

In \fref{fig3} one can see these peaks in the results of numerical simulation. Noteworthy, these peaks have a well pronounced triangular shape in the linear--log scale, meaning equation~\eref{eq-08} to represent the behaviour of the phase diffusion remarkably well for the major part of each multistability domain, not only at its centre. In \fref{fig4} one can see how the noise strength influences peaks in multistability domains. The numerical simulation data in all the figures are calculated with time series of length $2.5\cdot10^9$.

\section{Conclusion}
In this paper we have developed the theory of the effect of delayed feedback on coherence of noisy phase oscillators in the presence of frequency multistability induced by this time delay. The coherence has been quantified by the phase diffusion constant $D$. The process of alternation between two states has been demonstrated to be well representable as an asymmetric Markovian `telegraph' process. Employment of this `telegraphness' property allows constructing the framework for analytical study on the problem. The behaviour of the phase diffusion constant has been revealed to be smooth on the edges of the multistability domains. Giant peaks in the dependence of the phase diffusion constant on delay time have been discovered near points where the mean residence times in two states are equal (see equation~\eref{eq-08} and figures~\ref{fig3} and \ref{fig4}). Remarkably, their `integral strength' increases for vanishing noise, as their width $\propto\varepsilon^2$ while height is exponentially large in $1/\varepsilon^2$. For the longer delay time the peaks become taller.

\ack{
Authors acknowledge financial support by the Government of Perm Region (Contract {C-26/0004.3}) and the Russian Foundation for Basic Research (project no.\ 14-01-31380\_mol\_a).}

\end{document}